\documentclass [12pt]{article}
\usepackage{graphicx}
\usepackage[latin1]{inputenc}

\newcommand{\be}{\begin{equation}}
\newcommand{\ee}{\end{equation}}
\def\ba{\begin{eqnarray}}
\def\ea{\end{eqnarray}}
\def\ni{\noindent}
\def\bib{\bibitem}
\def\del{\partial}

\begin{document}
\baselineskip .75cm

\begin{center}
{\bf \Large Thermodynamics of Quasi-Particles at Finite Chemical
Potential}
\end{center}

\vspace{0.25 in}
\begin{center}
{F. G. Gardim$^{a}$ and  F. M. Steffens $^{a,b}$\\}
\vspace{0.25 in}
{\it \small \em $^{a}$ Instituto de F\'isica Te\'{o}rica -
Universidade
Estadual Paulista,\\
Rua Pamplona 145, 01405-900, S\~ao Paulo, SP, Brazil.\\}

\vspace{0.5cm}

{\it \small \em $^{b}$ NFC - CCH - Universidade
Presbiteriana Mackenzie,\\
Rua da Consola\c{c}\~ao 930, 01302-907, S\~ao Paulo, SP,
Brazil.\\}

\end{center}
\begin{abstract}

We present in this work a generalization of the solution of
Gorenstein and Yang to the inconsistency problem of thermodynamics
for systems of quasi-particles whose masses depend on both the
temperature and the chemical potential. We work out several
solutions for an interacting system of quarks and gluons and show
that there is only one type of solution that reproduce both
perturbative and lattice QCD.
\end{abstract}
\vspace{1cm}


\newpage

\section{Introduction}

In the past two decades, strongly coupled matter under extreme
conditions has been object of intensive study, as it must describe
from the core of some neutron stars to important features of the
early universe and heavy ion collisions experiments (see e.g.
\cite{Wilczek,karsch1}). It is the understanding that hadrons are
composed by asymptotically free particles that led to the idea that
at sufficiently high temperature $T$, and/or quark chemical
potential $\mu$, the hadronic matter should be described by quarks
and gluons degrees of freedom \cite{collins}, the quark-gluon plasma
(QGP). Current lattice calculations strongly suggests that this
phase transition actually happens, in the case of a vanishing
chemical potential, at a temperature around $T_c\simeq 190$MeV
\cite{Cheng:2006qk}. The search for a phase transition in the case
of a non-vanishing quark chemical potential has recently advanced,
as a result of new techniques for lattice calculations
\cite{Allton:2005gk,Fodor:2002km,lombardo}.

Results from ultra-relativistic heavy ion collisions at RHIC
\cite{RHIC1,RHIC2,RHIC3,RHIC4}, on the other hand, indicate that
the QGP has already been produced, and that hydrodynamics is
capable to describe the experimental data \cite{Shuryak:2003xe}.
The basic input for hydrodynamics is the equation of state EoS,
which must cover both the hadronic and the QGP sectors. The matter
created in the heavy ion collisions has high $T$ and small $\mu$,
therefore the study of the EoS at these conditions is of extreme
importance for the hydrodynamics approach. The
EoS of the hadronic phase is currently described by the hadronic
resonance gas model \cite{hagedorn}, with a good agreement to
lattice data \cite{Allton:2005gk,Karsch:2003vd}. For the
deconfined phase, one needs an EoS in terms of quark-gluon degrees
of freedom. QCD at finite $T$ and $\mu$ is, in principle, the
theoretical tool to compute the thermodynamics functions in this
new phase. However, strict perturbation theory, which have been
pushed to $g^6\ln(1/g)$ \cite{kajantie}, is reasonable only for
extremely high temperatures; at temperatures near $T_c$ it seems
not to be applicable, and alternative treatments appear to be
necessary. Moreover, the perturbative series seems to be weakly
convergent \cite{kajantie,arnold,zhai}. Specifically, it is
expected that when $T\gg T_c$, the plasma behaves like an ideal
gas of quarks and gluons, but the perturbative series converges at
a slow pace towards this expectation. A way to circumvent this
problem is through the reorganization of the perturbative series.
For instance, there are attempts to reorganize the series using
the Hard-Thermal-Loop (HTL) effective action \cite{braaten}, in
the form of so-called HTL perturbation theory
\cite{andersen,andersen2}, and also attempts based on the 2-loop
$\Phi$-derivable approximation \cite{blaizot,blaizot2}. The latter
approach assumes a massive quasi-particle formalism
\cite{pe.1,levai}.

Concomitantly, phenomenological models based on quasi-particles have
had success in describing the lattice data
\cite{pe.1,levai,peshier,s.1,rebhanroma,Bannur:2006hp,bannur,Gardim:2007ta},
over a wide range of temperatures (and at $\mu=0$), in addition to
being compatible with the fluid-like behavior found at RHIC
\cite{Peshier:2005pp}. More recently, a quasi-particle model for the
quark-gluon plasma (qQGP) for a non-vanishing $\mu$ has also been
worked out \cite{Bluhm:2007nu,Bannur:2007tk,Bluhm:2008sc}, with a
satisfactory description of the lattice data. Besides the
quasi-particle approach, there are other treatments to describe
lattice data near the transition region as, for instance, the
Polyakov loop enhanced Nambu-Jona-Lasinio (PNJL) model
\cite{Ghosh:2007wy,Roessner:2006xn}, which is able to fit lattice
data quite well, even when $\mu\neq 0$.

The quasi-particle model of the quark-gluon plasma is a
phenomenological model that assumes non-interacting massive
quasi-particles which, with the aid of few parameters, like the
thermal masses, is able to fit lattice QCD data over a wide range of
temperatures \cite{peshier,s.1,rebhanroma}: Not only at extremely
high temperatures, as in strict perturbation theory, or when
$T>20T_c$, as in the HTL perturbation theory, but also near $T_c$.
Indeed, quasi-particle models intend to describe deconfined matter
from $T_c$ up to $T\rightarrow\infty$, although from the point of
view of linking the QGP with hydrodynamics calculations, the
reproduction of lattice data near $T_C$ is somewhat more important.

The problem with the quasi-particle models is that the
thermodynamics relations calculated from them are not consistent
\cite{go.1}. Its origin is in the use of Standard Statistical
Mechanics (SSM) \cite{huang} with dispersion relations for
quasi-particles that resembles the dispersion relations for free
particles. Contrary to the case of particles, quasi-particles have
masses that are dependent on $T$ and/or $\mu$. The first work to
systematically study the consistency problem, and solve it for
finite $T$ but zero chemical potential, was the one by Gorenstein
and Yang \cite{go.1}. In their approach, it is built an effective
Hamiltonian which is composed of a part representing an ideal gas
and a part representing contributions from the vacuum. This extra
vacuum term leads to a modification in the pressure and in the
internal energy of the system, but keeps the entropy and the number
density untouched. In this way, a consistent thermodynamics for
quasi-particles is obtained. Most of the works in the literature are
based on this approach \cite{romache}. More recently, several
authors have also turned their attention to the thermodynamics
consistency of quasi-particle systems. In \cite{Biro:2001ug}, the
authors build an effective Hamiltonian, $H_{eff}$, as before but
require that the average of $\partial H_{eff}/\partial\mu$ and
$\partial H_{eff}/\partial T$ are zero. As a result, they get
expressions for the thermodynamics functions of the same kind found
in \cite{go.1}. On the other hand, in \cite{Yin:2007ns}, the
particle mass is taken as an independent variable. In this case, it
is also found a set of consistent thermodynamics relations. Other
possibility is presented in \cite{Bannur:2006hp}, where instead of
keeping the entropy unchanged, as Gorenstein and Yang, the internal
energy preserves its SSM definition, in a way that the consistency
of the thermodynamics relations is also achieved. Finally, a set of
solutions for the thermodynamics of quasi-particles at $\mu=0$ was
presented in \cite{Gardim:2007ta}. There it is shown that the
solutions of references \cite{Bannur:2006hp,go.1} are particular
solutions of a general formulation. The next natural step is to
extend this formulation to the case of finite $\mu$.

In this work, a framework for the thermodynamics of a system
composed of quasi-particles whose masses are $T$ and $\mu$ dependent
is developed, generalizing \cite{Gardim:2007ta}. At first sight, a
system composed of a mixture of fermions and bosons should follow
the law of Dalton for gas mixtures. This law states that {\it the
pressure exerted by a mixture of gases is equal to the sum of the
partial pressures of all the components present in the
mixture}\footnote{This stretch is a quotation from E. Fermi in
\cite{Fermi}.}. This statement holds exactly for ideal gases. For
real gases, however, the equality is approximated. As the
quasi-particle approach is built on the ideal gas model, one would
expect that the pressure for the whole system, i.e. for quarks and
gluons, were the sum of the fermionic and bosonic parts, separately.
In reality, this should not be true in the qQGP because the
underlining theory is that of an interacting system. It makes no
sense to fix the number of gluons $n_g=\del P_g/\del\mu=0$, since no
energy is expended for their creation ($\mu_g=0$). On the other hand
if the pressure is computed from Ref. \cite{Gardim:2007ta}, with a
suitable Debye mass, $m_D\equiv m_D(T,\mu)$, and including fermions,
it can be seen from the thermodynamics relations that the {\it
number of gluons} will be different from zero $n_g=\del
P_g(m_D)/\del\mu\neq 0$. Therefore, the law of Dalton fails here,
implying that it is necessary to develop the theory for the whole
system, and not for each part independently.

This article is organized as follows. In Sec. II, the general
formalism for a consistent thermodynamics of quasi-particles in the
grand canonical ensemble is presented. In sections III and IV,
particular solutions for this thermodynamics are computed, and in
section V a comparison with perturbative QCD is made. In section VI,
the comparison is with lattice QCD and, finally, the conclusions are
presented in section VII.

\section{The Formalism}

In the quasi-particle approach for the quark-gluon plasma, the
system is described by an effective Hamiltonian with mass terms that
are dependent on $T$ and/or $\mu$
\cite{pe.1,levai,Gardim:2007ta,Bannur:2007tk}. As widely known,
early attempts to implement this idea run in a serious problem: an
inconsistency in the thermodynamics relations. As shown by
Gorenstein and Yang \cite{go.1}, it is not enough to use the
Hamiltonian of an ideal gas with the mass $m$ replaced by the
effective mass $m(T,\mu)$. An extra term, $B$, has to be added to
the original Hamiltonian: $H \rightarrow H+\eta B$, where $\eta B$
is regarded as the zero point energy of the system. The extra term
modifies the usual thermodynamics functions, healing the
thermodynamics inconsistency of the system. Thus, the most general
form for the thermodynamics functions can be written as:

\begin{eqnarray}
\Phi(V,T,\mu,m_f^2,m_b^2)&=&\Phi_f(V,T,\mu,m_f^2)+\Phi_b(V,T,\mu,m_b^2)+{\alpha}B(V,m_f^2,m_b^2),\label{Phi}\\
U(V,T,\mu,m_f^2,m_b^2)&=&U_f(V,T,\mu,m_f^2)+U_b(V,T,\mu,m_b^2)+\eta B(V,m_f^2,m_b^2),\label{U}\\
S(V,T,\mu,m_f^2,m_b^2)&=&S_f(V,T,\mu,m_f^2)+S_b(V,T,\mu,m_b^2)-\frac{\gamma}{T}B(V,m_f^2,m_b^2),\label{S}\\
N(V,T,\mu,m_f^2,m_b^2)&=&N_{\{f\}}(V,T,\mu,m_f^2)-\frac{\lambda}{\mu}B(V,m_f^2,m_b^2),\label{N}
\end{eqnarray}
where $\Phi=-PV$ is the grand potential, $U$ is the internal energy,
$S$ is the entropy, and $N$ is the average net quark number. The
subscripts $f$ and $b$ denote fermions and bosons, respectively. The
functions $m_f^2 = m_f^2(T,\mu)$ and $m_b^2=m_b^2(T,\mu)$ are the
dynamical masses acquired by fermions and bosons. The constants
$\alpha$, $\eta$, $\gamma$ and $\lambda$ are arbitraries, but
constrained by $\alpha=\eta+\gamma+\lambda$, to guarantee the
consistency of the fourth thermodynamic relation in Eqs.
(\ref{Phi-rel}) below. This condition is just the extension of the
constraint obtained in Ref. \cite{Gardim:2007ta}. The $T$ and $\mu$
dependency of $B(V,m_f^2,m_b^2)$ is assumed to be through the masses
only. The extra term $B(V,m_f^2,m_b^2)$ must be such that the
thermodynamics relations

\begin{eqnarray}
P=-\left(\frac{{\del}\Phi}{{\del}V}\right)_{T,\mu},\hspace{.4cm}S=-\left(\frac{\del\Phi}{{\del}T}\right)_{V,\mu},
\hspace{.4cm}N=-\left(\frac{\del\Phi}{\del\mu}\right)_{T,V},\hspace{.4cm}U=\Phi+TS+{\mu}N.
\label{Phi-rel}
\end{eqnarray}
are valid. Substituting the Eqs. (\ref{Phi}), (\ref{U}), (\ref{S})
and (\ref{N}) in Eqs. (\ref{Phi-rel}), one finds the following pair
of differential equations:

\begin{eqnarray}
\gamma{B}=T\alpha\frac{\del{B}}{\del{T}}+T\sum_{i=b,f}\Bigg\langle\frac{\del{H_i}}{\del{T}}\Bigg\rangle,\hspace{1cm}
\lambda{B}=\mu\alpha\frac{\del{B}}{\del\mu}+\mu\sum_{i=b,f}\Bigg\langle\frac{{\del}H_i}{\del\mu}\Bigg\rangle,
\label{B.e.d-nao}
\end{eqnarray}
Using the connection between SSM and thermodynamics, $\Phi_i=-T\ln
Z_i$, where $Z_i$ is the grand partition function with Hamiltonian
$H_i$, one can deduce the following equalities:
$\langle\frac{\del{H_i}}{\del{T}}\rangle=\frac{{\del}m^2_i}{{\del}T}(\frac{\del\Phi_i}{\del{m^2_i}})_{T,V,\mu}$
and
$\langle\frac{\del{H_i}}{\del{\mu}}\rangle=\frac{{\del}m^2_i}{{\del}\mu}(\frac{\del\Phi_i}{\del{m^2_i}})_{T,V,\mu}$.
Using these expressions, and assuming $\alpha\neq 0$, one concludes
that the two equations in (\ref{B.e.d-nao}) are the same. They are
rewritten as:

\begin{eqnarray}
\frac{\del}{\del{m^2_i}}(BT^{-\frac{\gamma}{\alpha}}\mu^{-\frac{\lambda}{\alpha}})=
-\frac{T^{-\frac{\gamma}{\alpha}}\mu^{-\frac{\lambda}{\alpha}}}{\alpha}
\Bigg\langle\frac{\del{H_i}}{\del{m^2_i}}\Bigg\rangle
\hspace{1cm}\alpha\neq 0, \hspace{1cm} i=b,f.
\label{B.ed.alpha}
\end{eqnarray}
For $\alpha=0$, the limit $\alpha\rightarrow 0$ in Eqs.
(\ref{B.e.d-nao}) has to be taken with care. One obvious point is
that the two equations for $B$ in (\ref{B.e.d-nao}) give different
results if one simply sets $\alpha=0$\footnote{There is one trivial
possibility where B is the same, when $\gamma=\lambda$ and
$m^2_i=C_iT\mu$. But this is not the kind of solution that is
expected for the quasi-particle mass.}. When the grand potential has
no extra term, the average number of particles and the entropy must
have extra terms, i.e. if $\alpha=0$ then $\gamma$ and $\lambda$
must be different from zero. If such condition is not fulfilled than
the thermodynamics functions will not be consistent. This conclusion
is based on the fact that the entropy and the average number of
particles are obtained from the grand potential by Eqs.
(\ref{Phi-rel}). Therefore the previous redefined thermodynamics
functions are not valid when $\alpha = 0$, and the modified
thermodynamics functions must have the following form:

\begin{eqnarray}
\Phi=\Phi_b+\Phi_f,S=S_b+S_f-\gamma\frac{B_S}{T},N=N_{\{f\}}-\lambda\frac{B_N}{\mu},
U=U_b+U_f-\lambda{B_N}-\gamma{B_S}, \label{Phi-rel-0}
\end{eqnarray}
where $B_S$ and $B_N$ are

\begin{eqnarray}
B_S=\frac{T}{\gamma}\sum_{i=b,f}\Bigg\langle\frac{\del{H_i}}{\del{T}}\Bigg\rangle,\hspace{.5cm}
B_N=\frac{\mu}{\lambda}\sum_{i=b,f}\Bigg\langle\frac{\del{H_i}}{\del{\mu}}\Bigg\rangle\hspace{2.3cm}\alpha=0.
\label{B.sn}
\end{eqnarray}
The solutions for $B$ given by Eq. (\ref{B.sn}) are inversely
proportional to $\lambda$ and $\gamma$, respectively. Thus, Eqs.
(\ref{Phi-rel-0}) are independent of $\gamma$ and $\lambda$,
differently from the $\alpha\neq 0$ case.

As in the canonical case \cite{Gardim:2007ta}, the physical
meaning of $B(V,T,\mu)$ can be extracted from quantum statistical
mechanics. Using the quantum form for the internal energy, one has

\begin{eqnarray}
U=\langle\sum_{b,f}\hat{H}\rangle+{\eta}B=Tr[\hat{\rho}(\sum_{b,f}\hat{H}+\eta\hat{B})],
\label{U'}
\end{eqnarray}

\ni where $\hat{\rho}=e^{-\beta\hat{H}+\beta\mu\hat{N}}/Z$ is the
ensemble density operator, $\hat{H}$ is the Hamiltonian operator,
$\hat{N}$ is the number operator, and $\hat{B}\equiv B\textbf{1}$ is
the extra term with the unity operator. Note that when a term
independent of $q$ and $p$, but dependent on $T/\mu$, is added to
the Hamiltonian, the density operator for this modified Hamiltonian
is exactly the same as that obtained using the original Hamiltonian,
i.e. the density operator is defined up to a $T/\mu$ function in the
Hamiltonian. Thus, one can add $\hat{B}$ to $\hat{H}$ in
$\hat{\rho}$ without inducing any change in Eq. (\ref{U'}). Using
this property, it is seen that Eq. (\ref{U'}) has the same
definition as in the $SSM$ case: $U$ is given by the average value
of the system Hamiltonian. Therefore, the complete Hamiltonian,
particles plus vacuum, which must be used in the quasi-particle
picture, is $\hat{H}_T=\sum\hat{H}+\hat{E}_0$, with
$\hat{E}_0=\eta\hat{B}$ the zero point energy. The standard
procedure of discarding  the zero point energy can not be used here,
since it results in uncompleted thermodynamics functions, causing an
inconsistency.

For the average number of particles, the same manipulation (as for
the internal energy) can be applied to the $N$ quantum expression,
where instead of adding a $T/\mu$ function to the Hamiltonian, one
adds it to the $\hat{N}$ operator. Thus, ${N}$ follows the same
definition as in the $SSM$: ${N}$ is given by the mean value of the
total number operator, where the total number operator is
$\hat{N}_T=\hat{N}+\hat{N}_0$, with
$\hat{N}_0=-\frac{\lambda}{\mu}\hat{B}$. The extra term
$-\lambda\hat{B}/\mu$ represents the number of particles associated
with the zero point energy, which means that one can regard the zero
point energy of the quasi-particle scenario, multiplied by suitable
constants, as a mechanism to add or subtract particles from the
system.

These redefinitions imply the following thermodynamics functions:

\begin{eqnarray}
S=-\langle\ln\hat{\rho}\rangle-\gamma\frac{B}{T},\hspace{.4cm}\Phi=-T{\ln}Z_T+\gamma{B},\hspace{.4cm}N=\langle\hat{N}_T\rangle,\hspace{.4cm}U=\langle\hat{H}_T\rangle.
\label{finalexp-SUA}
\end{eqnarray}

\ni where $Z_T$ is the partition function with $\hat{H}_T$ and
$\hat{N}_T$. Note that for $\gamma=0$, the statistical mechanics
definitions for the entropy, the grand potential, the number of
particles and the internal energy are the usual ones.

The interpretation for $B$ in the $\alpha=0$ case follows closely
that of the $\alpha\neq 0$ case. Using again the quantum version of
Eq. (\ref{Phi-rel-0}), the internal energy can be written as
\begin{eqnarray}
U=U_b+U_f-\lambda{B_N}-\gamma{B_S}=\textrm{Tr}[\hat{\rho}(\sum_{i=b,f}\hat{H_i}-\lambda\hat{B}_N-\gamma\hat{B}_S)],
\label{U0}
\end{eqnarray}
where $\hat{B}_j\equiv B_j\textbf{1}$ is the operator for $j=N,S$.
As before, $\hat{H}_T=\sum\hat{H}+\hat{E}_0$, where
$\hat{E}_0=-\lambda\hat{B}_N-\gamma\hat{B}_S$ is the zero point
energy, dependent on $T$ and $\mu$ dependent. In the quantum
statistical mechanics picture, the self-consistent definition for
the number of particles is

\begin{eqnarray}
N=\langle\hat{N}\rangle-\frac{\lambda}{\mu}B_N=
\textrm{Tr}\left[\hat{\rho}\left(\hat{N}-\frac{\lambda}{\mu}\hat{B}_N\right)\right].
\label{N'0}
\end{eqnarray}
Applying the same arguments as before, the total number of
particles operator is $\hat{N}_T=\hat{N}+\hat{N}_0$, with
$\hat{N}_0=-\frac{\lambda}{\mu}\hat{B}_N$ the number of particles
associated with the zero point energy. Finally, the quantum
statistical mechanics relations are written as a function of the
total Hamiltonian

\begin{eqnarray}
\Phi=-T{\ln}Z_T+{\gamma}B_S,\hspace{.3cm}S=-\langle\ln\hat{\rho}\rangle-\gamma\frac{B_S}{T},\hspace{.3cm}N=\langle\hat{N}_T\rangle,\hspace{.3cm}U=\langle\hat{H}_T\rangle.
\label{finalexp-NUAS}
\end{eqnarray}
To conclude this section, the solutions for both $\alpha\neq 0$ and
$\alpha=0$ cases are summarized in the following table:
%
%

%

\section{Thermodynamics Functions for $\alpha\neq 0$}

To investigate the possible solutions for the thermodynamics of the
system, one needs an expression for the extra term $B$. From Eq.
(\ref{B.ed.alpha}), it is written as:

\begin{eqnarray}
{\alpha}B=B_0\mu^{\frac{\lambda}{\alpha}}T^{\frac{\gamma}{\alpha}}
-\mu^{\frac{\lambda}{\alpha}}T^{\frac{\gamma}{\alpha}}\sum_{i=b,f}{\int}dm^2_i\Bigg\langle\frac{\del{H_i}}{{\del}m_i^2}\Bigg\rangle\mu^{-\frac{\lambda}{\alpha}}T^{-\frac{\gamma}{\alpha}},\nonumber
\end{eqnarray}
or simply

\begin{eqnarray}
{\alpha}B=B_0\mu^{\frac{\lambda}{\alpha}}T^{\frac{\gamma}{\alpha}}
-\mu^{\frac{\lambda}{\alpha}}T^{\frac{\gamma}{\alpha}}\sum_{i=b,f}\left[{\int}d\mu\Bigg\langle\frac{\del{H_i}}{{\del}\mu}\Bigg\rangle\mu^{-\frac{\lambda}{\alpha}}T^{-\frac{\gamma}{\alpha}}+{\int}dT\Bigg\langle\frac{\del{H_i}}{{\del}T}\Bigg\rangle\mu^{-\frac{\lambda}{\alpha}}T^{-\frac{\gamma}{\alpha}}\right].
\label{B-1}
\end{eqnarray}
As the approach developed here is completely general, particular
choices for the constants $\alpha$, $\gamma$, $\lambda$, and $\eta$,
should reproduce known results in the literature, as for instance
the equations obtained by Gorenstein-Yang \cite{go.1}, and the ones
obtained by Bannur \cite{Bannur:2007tk}.

\subsection{The Solution of Gorenstein-Yang}

In Ref. \cite{go.1}, the thermodynamics of a system whose masses
depend on $T/\mu$ was studied through two conditions: (\textit{i})
the thermodynamics functions are defined as in the standard
Statistical Mechanics, and (\textit{ii}) the pressure is minimized
with respect to the phenomenological parameters. In the general
solution presented here, the condition $\gamma=0$ guarantees
(\textit{i}), implying that the vacuum does not contribute to the
total entropy of the system. The corresponding thermodynamics
functions are then given by:

\begin{eqnarray}
S&=&-\sum_{i=b,f}\langle \ln\hat{\rho_i}\rangle\nonumber\\
\Phi&=&-\sum_{i=b,f}\left(T{\ln}Z_i+\mu^{\frac{\lambda}{\alpha}}{\int}d\mu\Bigg\langle\frac{\del{H_i}}{\del\mu}\Bigg\rangle\mu^{-\frac{\lambda}{\alpha}}+\int{dT}\Bigg\langle\frac{\del H_i}{\del T}\Bigg\rangle\right)+{\alpha}B_0\mu^{\frac{\lambda}{\alpha}} \nonumber\\
N&=&\langle\hat{N}\rangle+\frac{\lambda}{\alpha}\mu^{-1}\sum_{i=b,f}\left(\int{dT}\Bigg\langle\frac{\del{H_i}}{\del{T}}\Bigg\rangle+\mu^{\frac{\lambda}{\alpha}}\int{d}\mu\Bigg\langle\frac{\del{H_i}}{\del\mu}\Bigg\rangle\mu^{-\frac{\lambda}{\alpha}}\right)-B_0\lambda\mu^{\frac{\lambda}{\alpha}-1}\nonumber\\
U&=&\sum_{i=b,f}\left[\langle\hat{H_i}\rangle-\frac{\eta}{\alpha}\left(\mu^{\frac{\lambda}{\alpha}}\int{d}\mu\Bigg\langle
\frac{\del{H_i}}{\del\mu}\Bigg\rangle\mu^{-\frac{\lambda}{\alpha}}+\int{dT}\Bigg\langle
\frac{\del{H_i}}{\del{T}}\Bigg\rangle\right)\right]+{\eta}B_0\mu^{\frac{\lambda}{\alpha}}.
\end{eqnarray}

The solutions with $\gamma = 0$ will be classified as of
Gorenstein-Yang type of solution. They can be divided into two
different branches.

\subsubsection{The Original Solution}

This is the most used solution, which satisfies both conditions
(\textit{i}) and (\textit{ii}), and was used in several studies of
QGP by quasi-particle model, as for instance in Refs.
\cite{pe.1,levai,peshier,s.1,rebhanroma}. In this approach, neither
the entropy nor the average number of particles are modified by any
additional term coming from the vacuum. This solution is obtained by
setting  $\gamma=\lambda=0$, and $\alpha=\eta$. The thermodynamics
functions are then reduced to:

\begin{eqnarray}
S&=&-\sum_{i=b,f}\langle \ln\hat{\rho_i}\rangle\nonumber\\
N&=&\langle\hat{N}\rangle\nonumber\\
\Phi&=&-\sum_{i=b,f}\left(T{\ln}Z_i+{\int}d\mu\Bigg\langle\frac{\del{H_i}}{\del\mu}\Bigg\rangle+\int{dT}\Bigg\langle\frac{\del H_i}{\del T}\Bigg\rangle\right)+B_0 \nonumber\\
U&=&\sum_{i=b,f}\left[\langle\hat{H_i}\rangle-\mu^{\frac{\lambda}{\alpha}}\int{d}\mu\Bigg\langle\frac{\del{H_i}}{\del\mu}
\Bigg\rangle\mu^{-\frac{\lambda}{\alpha}}-\int{dT}\Bigg\langle\frac{\del{H_i}}{\del{T}}\Bigg\rangle\right]+B_0.
\label{Bsol1A}
\end{eqnarray}
As this solution is well know and has been largely studied, we will
not provide here further details of its features, which can be found
in Ref. \cite{go.1}. It will be referred from now on as GY1.

\subsubsection{The Modified Solution}

A second solution of the type Gorenstein-Yang, is the one where
$\alpha=\lambda$ and $\gamma=\eta=0$, meaning that both the entropy
and the internal energy retain their original forms, while the other
2 thermodynamics functions are modified. Including bosons and
fermions, the thermodynamics functions are now written as:

\begin{eqnarray}
S&=&-\sum_{i=b,f}\langle \ln\hat{\rho_i}\rangle\nonumber\\
U&=&\sum_{i=b,f}\langle\hat{H_i}\rangle\nonumber\\
\Phi&=&-\sum_{i=b,f}\left(T{\ln}Z_i+\mu{\int}d\mu\Bigg\langle\frac{\del{H_i}}{\del\mu}\Bigg\rangle\mu^{-1}+\int{dT}\Bigg\langle\frac{\del H_i}{\del T}\Bigg\rangle\right)+B_0\mu \nonumber\\
N&=&\langle\hat{N}\rangle+\frac{\lambda}{\alpha}\mu^{-1}\sum_{i=b,f}\left(\int{dT}\Bigg\langle\frac{\del{H_i}}{\del{T}}
\Bigg\rangle+\mu\int{d}\mu\Bigg\langle\frac{\del{H_i}}{\del\mu}\Bigg\rangle\mu^{-1}\right)-B_0.
\label{Bsol1B}
\end{eqnarray}
This solution will be referred as GY2.

\subsection{The Solution for $\eta=\lambda=0$}

It is based on the principle that thermodynamics quantities that
have microscopic analogue must be computed by the average value of
its \textit{classic}\footnote{The word classic here means without
vacuum contributions.} counterpart. For example, the energy and
the average number of particles. Bannur studied this case for
$\mu=0$ in Ref. \cite{Bannur:2006hp},  and recently extended it to
$\mu\neq 0$ \cite{Bannur:2007tk}. In both cases, it was found good
agreement with lattice data, except in the region $T_c<T\leq
1.2T_c$. Setting $\eta=\lambda=0$ in Eqs. (\ref{finalexp-NUAS})
and (\ref{B-1}), one obtains:

\begin{eqnarray}
N&=&\langle\hat{N}\rangle \nonumber\\
U&=&\sum_{i=b,f}\langle\hat{H_i}\rangle \nonumber\\
S&=&-\sum_{i=b,f}\left(\langle\ln\hat{\rho_i}\rangle-\frac{1}{T}\int{d}\mu\Bigg\langle\frac{\del{H}_i}{\del\mu}\Bigg\rangle-\int{dT}\Bigg\langle\frac{\del{H_i}}{\del{T}}\Bigg\rangle{T}^{-1}\right)-B_0\nonumber\\
\Phi&=&-\sum_{i=b,f}\left(T\ln{Z_i}+\int{d}\mu\Bigg\langle\frac{\del{H}_i}{\del\mu}\Bigg\rangle+T\int{dT}\Bigg\langle
\frac{\del{H_i}}{\del{T}}\Bigg\rangle{T}^{-1}\right)+B_0T.
\label{Bsol2}
\end{eqnarray}

\subsection{Physical Constraint on $B$}

The extra term $B$, sometimes called bag energy or bag pressure,
plays the main role in the consistency of the thermodynamics
relations. Besides this, $B$ must also be physically acceptable. The
theory must guarantee that any physical solution containing $B$ must
recover the physics already known, as for instance reach the
Stefan-Boltzmann limit at high temperatures, or the perturbative
results of QCD at finite temperature and/or chemical potential. In
particular, in the limits where the masses are independent of $T$
and $\mu$, $B$ must be, at most, a constant:
\begin{eqnarray}
\lim_{\small{\begin{array}{cc}
  m_g\rightarrow 0\\
  m_q\rightarrow m_0 \\
\end{array}}}B(m_g,m_f,V)=B_0.
\end{eqnarray}
Using this condition in expression (\ref{B-1}), one obtains:

\begin{eqnarray}
\lim_{\small{\begin{array}{cc}
  m_g\rightarrow 0\\
  m_q\rightarrow m_0 \\
\end{array}}}B(m_g,m_f,V)=\alpha^{-1}B_0T^{\frac{\gamma}{\alpha}}\mu^{\frac{\lambda}{\alpha}}.
\end{eqnarray}
From this result it follows that $B_0$ must be zero if
$\gamma\neq 0$ and $\lambda\neq 0$.

Usually, $B$ is calculated from a differential equation deduced from
the Maxwell relation, $\del S/\del\mu=\del N/\del T$. This procedure
leads to an equation in the running coupling and in the extra term
$B$ \cite{rebhanroma,Bluhm:2007nu,romache}:

\begin{eqnarray}
a_T(T,\mu,g)\frac{\del g^2}{\del T}+a_{\mu}(T,\mu,g^2)\frac{\del
g^2}{\del\mu}=b(T,\mu,g).\nonumber
\end{eqnarray}
Thus, the knowledge of the coupling and $B$ at $\mu=0$, and at an
arbitrary $T$, is enough to determine these functions at any $T$ and
$\mu$. However, as done in this work, $B$ is given by the solution
of Eqs. (\ref{B.e.d-nao}). Naturally, one should expect both results
to give the same answer. In fact, they give. The general solution
presented here was built to be consistent. If one uses the full
expressions for the thermodynamics functions to verify whether the
Maxwell relation is respected or not, one finds that they indeed
respect it, without any additional condition on the masses or on the
coupling. Hence, one can do as in \cite{Bannur:2007tk}, and borrow
the masses of the particles (and the coupling) from any given
framework, without spoiling the consistency of the theory. The
choice of a given mass for the quasi-particles will induce a
particular form for $B$, leaving the whole theory self-consistent.
However, as will be shown here, this consistency holds only for the
full functions, and not in a perturbative expansion, with one
exception only.


\subsection{The Asymptotic Behavior of the Solutions}

The thermodynamics functions will be computed for a system
composed by quarks, anti-quarks and gluons, with a Hamiltonian
$H_i\equiv\omega_i=\sqrt{k^2+m^2_i(T,\mu)}$, where $|k|$ is the
magnitude of the quasi-particle momentum vector, and $m_i(T,\mu)$
is the quasi-particle mass. To obtain the asymptotic behavior of
the solutions, it is necessary to know the quasi-particle mass as
a function of $T$ and $\mu$. As statistical mechanics does not
provided this dependency, the standard procedure using the HTL
approach \cite{romache} for the computation of quasi-particles
masses will be adopted here. In next-to-leading order (NLO), the
mass for gluons and quarks are \cite{blaizot,blaizot2,rebhanroma}
:

\begin{eqnarray}
m_g^2&=&\frac{m_D^2}{2}-\frac{N_c}{\pi\sqrt{2}}g^2Tm_g,\nonumber\\
\widetilde{m}_q^2&=&\frac{N_gg^2}{8N_c}\left(T^2+\frac{\mu^2}{\pi^2}\right)-\frac{N_g}{N_c}\frac{\sqrt{2}g^2}{4\pi}Tm_g,
\label{masses}
\end{eqnarray}
where $N_g=(N_c^2-1)$, $N_c$ is the number of colors, $N_f$ is the
number of flavors, and
$m_D^2=g^2\left(\frac{2N_c+N_f}{6}T^2+\frac{N_f}{2}\frac{\mu^2}{\pi^2}\right)$
is the Debye mass \cite{blaizot,blaizot2,romache}. Including the
bare quark mass, $m_0$, one writes the quark mass as
$m_q=\sqrt{m^2_0+\widetilde{m}_q^2}$. For the numerical
calculations presented in this work, the value $m_0/T=0.4$ will be
used, according to Ref. \cite{Allton:2005gk}. Using the
approximations $g\ll 1$ (and $\mu/T\ll 1$), the quark mass is then
written as


\begin{eqnarray}
m_q^2=m^2_0+\frac{N_gg^2T^2}{8N_c}\left[1+\frac{\mu^2}{\pi^2T^2}-\frac{g}{\pi}\left(2\sqrt{\frac{2N_c+N_f}{6}}+\frac{3N_f}{\sqrt{12N_c+6N_f}}\frac{\mu^2}{\pi^2T^2}\right)+\frac{N_cg^2}{\pi^2}\right].
\label{mass}
\end{eqnarray}
The coupling constant $g$ will be replaced by an effective coupling
constant $g_s$ \cite{Bluhm:2007nu}, inspired by perturbative QCD, in
order to accommodate non-perturbative effects at the vicinity of
$T_c$ when fitting the lattice data. In that case,

\begin{eqnarray}
g^2_s(T,\mu)=\frac{16\pi^2}{\beta_0\ln\xi^2}\left[1-\frac{2\beta_1}{\beta_0^2}\frac{\ln(\ln\xi^2)}{\ln\xi^2}\right],
\label{g_s}
\end{eqnarray}
with $\beta_0=(11N_c-2N_f)/3$,
$\beta_1=(34N^2_c-13N_fN_c+3N_f/N_c)/6$, and
$\xi\equiv\lambda_s\frac{T}{T_c}\sqrt{1+u^2\frac{\mu^2}{T^2}}-\lambda_s\frac{T_s}{T_c}$,
where $\lambda_s$ is a scale parameter and $T_s$ is a temperature
shift that regulates the infrared divergence below the critical
temperature $T_c$. The factor $\sqrt{1+u^2\frac{\mu^2}{T^2}}$ in
$\xi$ comes from the study of QCD running coupling at finite
temperature and quark chemical potential based on semiclassical
background field method \cite{Schneider:2003uz}. The parameter $u$
will be chosen to provide a best fit for the lattice data. For
$\mu/T\rightarrow 0$, Eq. (\ref{g_s}) is reduced to the usual result
for the coupling constant.

The solutions for $\alpha\neq 0$ are not easy to calculate
explicitly since it is necessary to perform integrals on both $T$
and $\mu$ to find $B$. Nevertheless, in the limits
$T\rightarrow\infty$ and $\mu\rightarrow\infty$ calculations can be
done. Eq. (\ref{B-1}) is rewritten as:

\begin{eqnarray}
{B}=\frac{B_0}{\alpha}\mu^{\frac{\lambda}{\alpha}}T^{\frac{\gamma}{\alpha}}-\frac{\mu^{\frac{\lambda}{\alpha}}T^{\frac{\gamma}{\alpha}}}{\alpha}\sum_{b,f}\left[{\int}d\mu\Bigg\langle\frac{\del{H_i}}{{\del}\mu}\Bigg\rangle\mu^{-\frac{\lambda}{\alpha}}T^{-\frac{\gamma}{\alpha}}+{\int}dT\Bigg\langle\frac{\del{H_i}}{{\del}T}\Bigg\rangle\mu^{-\frac{\lambda}{\alpha}}T^{-\frac{\gamma}{\alpha}}\right]=B_b+B_f,
\label{B}
\end{eqnarray}
where $B_b$ is the piece correspondent to gluons, and $B_f$ contains
$B_0$ and the quark part. The average value of the derivatives of
the Hamiltonian (with respect to the masses) are computed, and the
results are:

\begin{eqnarray}
\Bigg\langle\frac{\del{H_b}}{{\del}m^2_b}\Bigg\rangle&=&\frac{Vd_b}{4\pi^2}\int{dk}\frac{k^2}{\sqrt{k^2+m^2_b}}n^+_b(k)=\frac{d_bVT^2}{2\pi^2}J_3\left(\frac{m_b}{T}\right)\approx\frac{d_bVT^2}{24};
\label{delHb}\\
\Bigg\langle\frac{\del{H_f}}{{\del}m^2_f}\Bigg\rangle&=&
\frac{Vd_f}{4\pi^2}\int{dk}\frac{k^2}{\sqrt{k^2+m^2_f}}n^+_f=
\frac{d_fVT^2}{2\pi^2}I^T_3\left(\frac{m_f}{T},\frac{\mu}{T}\right)
\approx\frac{d_fVT^2}{2\pi^2}\left(\frac{\pi^2}{12}+\frac{\mu^2}{4T^2}\right),
\label{delHf}
\end{eqnarray}
where $d_b$ and $d_f$ are the degeneracy factors of bosons and
fermions, respectively. The functions $I^T_i$ and $J_i$ are given in
Appendix A. The derivatives of the mass with respect to $T$ and
$\mu$ are

\begin{eqnarray}
\frac{\del{m^2_b}}{\del{\mu}}=g^2\frac{N_f}{2}\frac{\mu}{\pi^2},&&\hspace{2cm}\frac{\del{m^2_b}}{\del{T}}=g^2\frac{2N_c+N_f}{6}T;
\label{delmb}\\
\frac{\del{m^2_f}}{\del{\mu}}=g^2\frac{N_g}{4N_c}\frac{\mu}{\pi^2},&&\hspace{2cm}\frac{\del{m^2_f}}{\del{T}}=g^2\frac{N_g}{4N_c}T.
\label{delmf}
\end{eqnarray}
Note that the asymptotic approximation implies that the masses are
calculated in leading order. With these four equations it is easy to
compute $B$. For the gluon contribution, using Eqs. (\ref{B}),
(\ref{delHb}), and (\ref{delmb}), one has

\begin{eqnarray}
\frac{B_b\alpha}{Vd_b}\approx-\frac{m^2_bT^2}{12(2-\Lambda)}+g^2\frac{(2N_c+N_f)T^4}{12^2}\frac{2+\Lambda-\Gamma}{(2-\Lambda)(4-\Gamma)},
\label{Bb}
\end{eqnarray}
where $\Lambda=\frac{\lambda}{\alpha}$ and
$\Gamma=\frac{\gamma}{\alpha}$. From Eqs. (\ref{B}), (\ref{delHf})
and (\ref{delmf}), the quark and anti-quark contribution is written
as

\begin{eqnarray}
\frac{B_f\alpha}{Vd_f} {\approx} B_0\mu^{\Lambda}T^{\Gamma}-\frac{m^2_f}{12(4-\Gamma)}\left(T^2+\frac{3\mu^2}{\pi^2}\right)-\frac{g^2N_g\mu^2}{32\pi^2(4-\Gamma)N_c}\left[\frac{\Lambda-\Gamma}{4-\Lambda}\frac{\mu^2}{\pi^2}+\left(\frac{2}{2-\Gamma}+\right.\right.\nonumber\\
\left.\left.\hspace{2cm}+\frac{2-\Gamma+\Lambda}{3(2-\Lambda)}\right)T^2\right].
\label{Bf}
\end{eqnarray}
In the calculation of $B_b$ and $B_f$ at high $T$ and $\mu$, the
following asymptotic conditions were used: $\frac{T_0}{T}\ll 1$,
and $\frac{\mu_0}{\mu}\ll 1$, where $T_0$ and $\mu_0$ are the
lower limits of the integrals. Also, in the region $g\rightarrow
0$, $g$ is a very slowly function of $T$ and $\mu$, what justifies
somewhat to regard $g$ in the mass derivatives as constants, Eqs.
(\ref{delmb}) and (\ref{delmf}). The integrals converge in this
region only if conditions $\Lambda<2$ and $\Gamma<2$ are
satisfied. Inspired by the case of a system made of bosons only at
$\mu=0$ \cite{Gardim:2007ta}, this condition will be called of
{\it weak physical condition}. These results provide the necessary
quantities to describe the thermodynamics of the system. To write
the asymptotic pressure one needs  Eqs. (\ref{Phi}), (\ref{Bb})
and (\ref{Bf}). Then,

\begin{eqnarray}
P(T,\mu)=\frac{\pi^2}{90}\left(\frac{7N_cN_f+4N_g}{2}T^4+15N_cN_f\frac{\mu^2T^2}{\pi^2}+\frac{15N_cN_f}{2}\frac{\mu^4}{\pi^4}\right)+\hspace{2cm}\nonumber\\
-\frac{N_gg^2}{32}\left(\frac{4N_c+5N_f}{9}\frac{2-\Gamma}{4-\Gamma}T^4+2N_f\frac{\Gamma\Lambda-\Gamma-\Lambda}
{(2-\Gamma)(2-\Lambda)}\frac{\mu^2T^2}{\pi^2}+N_f\frac{2-\Lambda}{4-\Lambda}\frac{\mu^4}{\pi^4}\right).
\label{Pt}
\end{eqnarray}
The term independent of $g$ is the Stefan-Boltzmann pressure,
while the rest of the expression is the correction to order $g^2$.
Using Eq. (\ref{Pt}) in the thermodynamics relations Eq.
(\ref{Phi-rel}) is possible to obtain the other expressions. On
the other hand, following the same steps used to calculate the
pressure, one can do the calculation of the other quantities
directly. As the theory has thermodynamic consistency, both ways
should give the same result. However, consistency has been proved
only for the full thermodynamics functions. One question then
arises: Are the perturbative thermodynamics functions also
consistent order by order? To answer this question is necessary to
compute $S$, $N$ and $U$ in both manners and compare the results.

The asymptotic entropy density, $s=S/V$, is calculated from Eqs.
(\ref{S}), (\ref{Bb}) and (\ref{Bf}):

\begin{eqnarray}
s(T,\mu)=\frac{\pi^2T}{3}\left(\frac{7N_cN_f+4N_g}{15}T^2+N_cN_f\frac{\mu^2}{\pi^2}\right)+\hspace{5.4cm}\nonumber\\
-\frac{N_gg^2T}{24}\left[\frac{4N_c+5N_f}{3}\frac{2-\Gamma}{4-\Gamma}T^2+\frac{3N_f}{2}\frac{4-6\Gamma-2\Lambda+2\Gamma\Lambda+\Gamma^2}{(2-\Gamma)(2-\Lambda)}\frac{\mu^2}{\pi^2}-\frac{3N_f\Gamma}{2(4-\Lambda)}\frac{\mu^4}{\pi^4T^2}\right].
\label{st}
\end{eqnarray}
For the asymptotic particle number density, $n=N/V$,  one writes
it in terms of the integral involving fermions, which can be found
in Appendix \ref{App-A}: $\langle N\rangle =
\frac{d_fT^3}{\pi^2}G_3^{T}\left(\frac{m_f}{T},\frac{\mu}{T}\right)$.
The net number density is then written as:

\begin{eqnarray}
n(T,\mu)=N_cN_f\frac{\mu}{3}\left(T^2+\frac{\mu^2}{\pi^2}\right)+\hspace{7.6cm}\nonumber\\
-\frac{N_gg^2\mu}{16\pi^2}\left(N_f\frac{4-2\Gamma-6\Lambda+2\Gamma\Lambda+\Lambda^2}{(2-\Gamma)(2-\Lambda)}T^2+\frac{2N_f(2-\Lambda)}{(4-\Lambda)}\frac{\mu^2}{\pi^2}+\frac{4N_c+5N_f}{9(4-\Gamma)}\frac{\Lambda\pi^2
T^4}{\mu^2}\right). \label{nt}
\end{eqnarray}
On the other hand, computing the entropy and the average number of
particles from the result for the pressure, Eq. (\ref{Pt}), using
the thermodynamics relations Eq. (\ref{Phi-rel}), one obtains:

\begin{eqnarray}
s_{therm}=\frac{{\del}P}{\del{T}}=\frac{\pi^2T}{3}\left(\frac{7N_cN_f+4N_g}{15}T^2+N_cN_f\frac{\mu^2}{\pi^2}\right)+\hspace{4cm}\nonumber\\
-\frac{N_gg^2T}{8}\left[\frac{4N_c+5N_f}{9}\frac{2-\Gamma}{4-\Gamma}T^2+N_f\frac{\Gamma\Lambda-\Gamma-\Lambda}{(2-\Gamma)(2-\Lambda)}\frac{\mu^2}{\pi^2}\right];\nonumber\\
n_{therm}=\frac{{\del}P}{\del{\mu}}=\frac{N_cN_f\mu}{3}\left(T^2+\frac{\mu^2}{\pi^2}\right)-\frac{N_gN_fg^2\mu}{8\pi^2}\left[\frac{\Gamma\Lambda-\Gamma-\Lambda}{(2-\Gamma)(2-\Lambda)}T^2+\frac{2-\Lambda}{4-\Lambda}\frac{\mu^2}{\pi^2}\right].
\label{nt-st-termo}
\end{eqnarray}

Trivially, the resulting calculation for the independent
quasi-particle terms, i.e. the terms independent of $g$, are
consistent: They result the same expression for both procedures.
However, the terms in $g^2$ fails the consistency check. Take, for
instance, the expressions for the entropy. Only the terms that
depend solely on the temperature match when using both procedures.
The same happens for the particle number density, only now instead
of the terms involving the temperature, are the terms involving
the chemical potential the ones that matches. The terms involving
both the temperature and the chemical potential do not match in
any situation. Comparing both procedures for the terms in $\mu
T^2$ of $n$ and in $T\mu^2$ of $s$, the unique solution which
produces a possible agreement between the two procedures is the
one with $\Gamma=\Lambda=2$. But this solution was excluded in the
$B$ calculation, the weak physical condition, because the
integrals do not converge. Concluding, the asymptotic limit of
thermodynamics functions breaks thermodynamics consistency order
by order in a perturbative expansion. Does the same happen with
the $\alpha=0$ solutions?


\section{The $\alpha=0$ solution}
The main feature of the $\alpha=0$ solution is the fact that the
extra terms are easier to compute, as they do not involve integrals
on $T$ and/or $\mu$. In such case, it is possible to have an
analytical result for almost all range of temperature and chemical
potential. In order to obtain the thermodynamics of the system, it
is convenient to start with the pressure, as there are no extra
terms contributing to it. The pressure of an ideal gas with modified
masses is

\begin{eqnarray}
P(T,\mu)=\frac{1}{6\pi^2}\int_0^{\infty}dkk^4\sum_{i=b,f}d_i\frac{n^+_i(k)}{\sqrt{k^2+m^2_i(T,\mu)}},
\label{Pideal}
\end{eqnarray}
where the distribution functions are
$n^\pm_f(k)=[e^{\beta(\sqrt{k^2+m_f^2}-\mu)}+1]^{-1}\pm[e^{\beta(\sqrt{k^2+m_f^2}+\mu)}+1]^{-1}$
and
$n^\pm_b(k)=\frac{1}{2}([e^{\beta(\sqrt{k^2+m_b^2})}-1]^{-1}\pm[e^{\beta(\sqrt{k^2+m_b^2})}-1]^{-1})$.
Expression (\ref{Pideal}) can be computed explicitly, see Appendix
\ref{App-A}, using convergent boundary for fermions:(i)
$\frac{\mu}{T},\frac{m_f}{T}<\pi$ or (ii)
$\Big|\frac{a^2}{\pi^2}+\frac{2ai}{\pi}-\frac{r_f^2}{\pi^2}\Big|<1$,
where $a=\frac{\mu}{T}$ and $r_f=\frac{m_f}{T}$. For bosons, the
boundary condition is $r_b=\frac{m_b}{T}<2\pi$. The pressure can
then be written as
$P=\frac{4}{\pi^2}T^4[d_fI^T_5(r_f,a)+d_bJ_5(r_b)]$, or

\begin{eqnarray}
P=\frac{\pi^2T^4}{90}\left\{d_b+\frac{7d_f}{4}+\frac{15a^2d_f}{2\pi^2}+\frac{15a^4d_f}{4\pi^4}-\frac{15(d_fr_f^2+d_br_b^2)}{4\pi^2}-\frac{45a^2r_f^2d_f}{4\pi^4}+\frac{15r^3_bd_b}{2\pi^2}+\right.\nonumber\\
+\frac{45}{8\pi^4}\left[r_f^4d_f\left(\frac{3}{4}-\gamma_E-\ln\frac{r_f}{\pi}\right)-\frac{r_b^4d_b}{2}\left(\frac{3}{4}-\gamma_E-\ln\frac{r_b}{4\pi}\right)\right]+\frac{105\zeta(3)d_f}{64\pi^6}(r_f^2-a^2)^3+\nonumber\\
\left.+\frac{45(r_f^2-a^2)^2d_f}{64\pi^2}\left[21\zeta(3)a^2-\frac{31\zeta(5)}{32\pi^2}(r_f^2-a^2)^2\right]+\frac{45d_b}{8\pi^4}\sum^{\infty}_{n=1}\frac{(-1)^n\Gamma(n+\frac{1}{2})r_b^{2(n+2)}}{(2\pi)^{2n}\Gamma\left(\frac{1}{2}\right)\Gamma(n+3)}+\right.\nonumber\\
\left.-d_f\sum_{n=0}^\infty\frac{45{(2n+1)^3}}{2}\sum_{k=2}^{\infty}\frac{\Gamma(\frac{1}{2}+k)}{\Gamma(\frac{1}{2})\Gamma(k+3)}\Re\left(\frac{a^2-r_f^2}{[(2n+1)\pi]^2}+\frac{2ai}{(2n+1)\pi}\right)^{k+2}\right\}.
\label{P(Tmu)}
\end{eqnarray}
For the computation of the entropy, Eqs. (\ref{Phi-rel-0}) and
(\ref{B.sn}) are necessary. In this case,

\begin{eqnarray}
s(T,\mu)&=&\frac{1}{2\pi^2T}\int_0^{\infty}dkk^2\sum_{i=b,f}d_i\left[\frac{\frac{4}{3}k^2+\left(m_i^2-\frac{T}{2}\frac{\del{m_i^2}}{\del{T}}\right)}{\sqrt{k^2+m_i^2}}n^+_i(k)-\mu n^-_i(k)\right]\nonumber\\
&=&\frac{d_fT^3}{\pi^2}\left[16I^T_5(r_f,a)+\left(\frac{m_f^2}{T^2}-\frac{1}{2T}\frac{\del{m_f^2}}{\del{T}}\right)I^T_3(r_f,a)-\frac{\mu}{T}G^T_3(r_f,a)\right]+\nonumber\\
&&\hspace{3cm}+\frac{d_bT^3}{\pi^2}\left[16J_5(r_b)+\left(\frac{m_b^2}{T^2}-\frac{1}{2T}\frac{\del{m_b^2}}{\del{T}}\right)J_3(r_b)\right].
\label{S(Tmu)}
\end{eqnarray}
An explicit integration can be done in this expression for $s$ using
the integrals from Appendix \ref{App-A}. The same sort of
manipulation can be made for the case of the internal energy
density, $e\equiv U/V$. The result is:

\begin{eqnarray}
e(T,\mu)=\frac{1}{2\pi^2}\int_0^{\infty}dk\sum_{i=b,f}\frac{k^2}{\sqrt{k^2+m_i^2}}\left[k^2+m_i^2-\frac{1}{2}\left(\mu\frac{\del{m_i^2}}{\del{\mu}}+T\frac{\del{m_i^2}}{\del{T}}\right)\right]d_in^+_i(k)\nonumber\\
=\frac{d_fT^4}{\pi^2}\left\{12I^T_5(r_f,a)+\left[\frac{m_f^2}{T^2}-\frac{1}{2T^2}\left(\mu\frac{\del{m_f^2}}{\del{\mu}}+T\frac{\del{m_f^2}}{\del{T}}\right)\right]I^T_3(r_f,a)\right\}+\nonumber\\
\hspace{1cm}+\frac{d_bT^4}{\pi^2}\left\{12J_5(r_b)+\left[\frac{m_b^2}{T^2}-\frac{1}{2T^2}\left(\mu\frac{\del{m_b^2}}{\del{\mu}}+T\frac{\del{m_b^2}}{\del{T}}\right)\right]J_3(r_b)\right\}.
\label{E(Tmu)}
\end{eqnarray}
Similarly, the result for the average number of particles is

\begin{eqnarray}
n(T,\mu)&=&\frac{1}{2\pi^2}\int_0^{\infty}dkk^2\sum_{i=b,f}d_i\left[n_i^-(k)-\frac{1}{2\sqrt{k^2+m_i^2}}\frac{\del{m_i^2}}{\del{\mu}}n_i^+(k)\right]\nonumber\\
&=&\frac{T^3}{\pi^2}\left\{d_fG^T_3(r_f,a)-\frac{1}{2T}\left[d_f\frac{\del{m_f^2}}{\del{\mu}}I^T_3(r_f,a)+d_b\frac{\del{m_b^2}}{\del{\mu}}J_3(r_b)\right]\right\}.
\label{N(Tmu)}
\end{eqnarray}
As done in the $\alpha\neq 0$ case, it is necessary investigate the
consistency of the expanded thermodynamics functions. The pressure
at leading order can be obtained from Eq. (\ref{P(Tmu)}),

\begin{eqnarray}
P(T,\mu)=\frac{\pi^2}{6}\left(\frac{7N_cN_g+4N_g}{30}T^4+N_fN_c\frac{\mu^2T^2}{\pi^2}+\frac{N_fN_c}{2}\frac{\mu^4}{\pi^4}\right)+\hspace{3cm}\nonumber\\
-\frac{N_gg^2}{16}\left(\frac{4N_c+5N_f}{18}T^4+N_f\frac{\mu^2T^2}{\pi^2}+\frac{N_f}{2}\frac{\mu^4}{\pi^4}\right).
\label{P0}
\end{eqnarray}
The leading order results for $s$ and $n$ are calculated from Eqs.
(\ref{S(Tmu)})and (\ref{N(Tmu)}). The results are:

\begin{eqnarray}
s(T,\mu)&=&\frac{\pi^2T}{3}\left(\frac{7N_cN_f+4N_g}{15}T^2+N_cN_f\frac{\mu^2}{\pi^2}\right)-\frac{g^2TN_g}{8}\left(\frac{4N_c+5N_f}{9}T^2+N_f\frac{\mu^2}{\pi^2}\right),\nonumber\\
n(T,\mu)&=&\frac{N_cN_f\mu}{3}\left(T^2+\frac{\mu^2}{\pi^2}\right)-\frac{g^2N_gN_f\mu}{8\pi^2}\left(T^2+\frac{\mu^2}{\pi^2}\right).
\label{ns0SM}
\end{eqnarray}

The entropy density and the particle number density can also be
calculated directly from the thermodynamics relations. As before, we
denote them by $s_{therm}$ and $n_{therm}$, respectively. Opposite
to the $\alpha\neq 0$ case, the resulting expressions now agree: For
$\alpha=0$, $s=s_{therm}$ and $n=n_{therm}$ at order $g^2$.


\section{Comparison with Perturbative QCD}

The quasi-particle picture has been introduced to describe the QGP
mainly because perturbative QCD is unable to study the QGP close to
the deconfinement region. Perturbative calculations, indeed, exhibit
a poor convergence except for temperatures as high as $>10^5T_c$. In
any case, perturbative QCD calculations are what the model should
reproduce at extremely high temperatures and chemical potential.

The thermodynamics functions, to order $g^2$, in perturbative
QCD are \cite{blaizot,blaizot2}:

\begin{eqnarray}
P_{QCD}=\frac{\pi^2}{6}\left(\frac{7N_cN_g+4N_g}{30}T^4+N_fN_c\frac{\mu^2T^2}{\pi^2}+\frac{N_fN_c}{2}\frac{\mu^4}{\pi^4}\right)+\hspace{3.8cm}\nonumber\\
-\frac{N_gg^2}{32}\left(\frac{4N_c+5N_f}{18}T^4+N_f\frac{\mu^2T^2}{\pi^2}+\frac{N_f}{2}\frac{\mu^4}{\pi^4}\right);\nonumber\\
s_{QCD}=\frac{\pi^2T}{3}\left(\frac{7N_cN_f+4N_g}{15}T^2+N_cN_f\frac{\mu^2}{\pi^2}\right)-\frac{g^2TN_g}{16}\left(\frac{4N_c+5N_f}{9}T^2+N_f\frac{\mu^2}{\pi^2}\right);\nonumber\\
n_{QCD}=\frac{N_cN_f\mu}{3}\left(T^2+\frac{\mu^2}{\pi^2}\right)-\frac{g^2N_gN_f\mu}{16\pi^2}\left(T^2+\frac{\mu^2}{\pi^2}\right).\hspace{5cm}
\label{psnQCD}
\end{eqnarray}
%


It is fundamental that the thermodynamics functions calculated here
be compared with the results from perturbative QCD. First, the
solutions for $\alpha\neq 0$ are compared. As discussed before,
these solutions have the serious problem of not being consistent
order by order. Nevertheless, the comparison with perturbative QCD
is made using expressions (\ref{Pt}), (\ref{st}), and (\ref{nt})
computed directly from their definitions. If one uses, instead, Eqs.
(\ref{nt-st-termo}), the expressions for the thermodynamics
functions would be different, but the final conclusion would be the
same. The cases discussed are:

\begin{enumerate}
    \item {$\gamma=\lambda=0$, $\alpha\neq 0$: Solution GY1.
\begin{eqnarray}
P_{GY1}=-\frac{g^2N_g}{64}\left(\frac{4N_c+5N_f}{9}T^4+N_f\frac{\mu^4}{\pi^4}\right),\hspace{3cm}\nonumber\\
s_{GY1}=-\frac{g^2N_gT}{16}\left(\frac{4N_c+5N_f}{9}T^2+N_f\frac{\mu^2}{\pi^2}\right),\hspace{1cm}n_{GY1}=-\frac{g^2N_gN_f\mu}{16\pi^2}\left(T^2+\frac{\mu^2}{\pi^2}\right)\nonumber
\end{eqnarray}
Although this is the solution widely used in the literature
\cite{pe.1,levai,peshier,s.1,rebhanroma}, the asymptotic expression
for the pressure does not agree with perturbative QCD, although $s$
and $n$ agree. However, this agreement is purely accidental given
the lack of consistency of the thermodynamics functions at this
order, as discussed in section 3.4. Note that the whole problem
arises in the mixed term, $\mu^2T^2$, in the pressure. This is the
first explicit calculation of these 3 functions.}
    \item {$\gamma=\eta=0$, $\alpha\neq 0$: Solution GY2.
\begin{eqnarray}
P_{GY2}&=&-\frac{g^2N_g}{32}\left(\frac{4N_c+5N_f}{18}T^4-\frac{N_f}{2}\frac{\mu^2T^2}{\pi^2}+\frac{N_f}{3}\frac{\mu^4}{\pi^4}\right),\nonumber\\
s_{GY2}&=&-\frac{g^22N_gT}{16}\left(\frac{4N_c+5N_f}{9}T^2+N_f\frac{\mu^2}{\pi^2}\right),\nonumber\\
n_{GY2}&=&-\frac{g^2N_g\mu}{16\pi^2}\left(\frac{4N_c+5N_f}{36}\frac{\pi^2
T^4}{\mu^2}-\frac{N_f}{2}T^2+\frac{2N_f}{3}\frac{\mu^2}{\pi^2}\right)\nonumber
\end{eqnarray}
Here only $s$ match with perturbative QCD.}
    \item {$\lambda=\eta=0$, $\alpha\neq 0$: Solution proposed by Bannur \cite{Bannur:2006hp,Bannur:2007tk}.
\begin{eqnarray}
P_{B}=-\frac{g^2N_g}{32}\left(\frac{4N_c+5N_f}{27}T^4-N_f\frac{\mu^2T^2}{\pi^2}+\frac{N_f}{2}\frac{\mu^4}{\pi^4}\right),\hspace{3cm}\nonumber\\
s_B=-\frac{g^2N_gT}{24}\left(\frac{4N_c+5N_f}{9}T^2-\frac{3N_f}{4}\frac{\mu^2}{\pi^2}-\frac{3N_f}{8}\frac{\mu^4}{\pi^4T^2}\right),\hspace{.3cm}n_B=-\frac{g^2N_gN_f\mu}{16\pi^2}\left(T^2+\frac{\mu^2}{\pi^2}\right).\nonumber
\end{eqnarray}
As expected, the only function whose asymptotic form agrees with QCD
is $n$.
    }
    \item {$\alpha = 0$: The consistent solution.
\begin{eqnarray}
P_{\alpha = 0}=-\frac{N_gg^2}{16}\left(\frac{4N_c+5N_f}{18}T^4+N_f\frac{\mu^2T^2}{\pi^2}+\frac{N_f}{2}\frac{\mu^4}{\pi^4}\right),\hspace{3cm}\nonumber\\
s_{\alpha=0}=-\frac{g^2TN_g}{8}\left(\frac{4N_c+5N_f}{9}T^2+N_f\frac{\mu^2}{\pi^2}\right),\hspace{.3cm}n_{\alpha=0}=-\frac{g^2N_gN_f\mu}{8\pi^2}\left(T^2+\frac{\mu^2}{\pi^2}\right).\nonumber
\end{eqnarray}
Unlike the 3 precedent cases, the solutions with $\alpha = 0$ are
the {\it only} solutions which are consistent order by order. By
extension, they are the only solutions that are meaningful when
expanding in $g^2$. However, when using the HTL/HDL masses for
the explicit calculation of the thermodynamics functions, Eqs.
(\ref{P0}) and (\ref{ns0SM}), one gets half the result of
perturbative QCD for the 3 functions.
    }

\end{enumerate}
The relevant question now is: Is there any scheme that is consistent
order by order in the asymptotic limit and that also reproduces
perturbative QCD? As seen, the only possibility of consistent
schemes at finite $T$ {\it and} $\mu$ are the ones with $\alpha=0$.
However, to match perturbative QCD, a quasi-mass for quarks and
gluons that is not the HTL/HDL mass has to be used. Inspired by the
calculation at finite $T$ and $\mu=0$ of Ref. \cite{Gardim:2007ta},
where the HTL mass was replaced by $m_{HTL}/\sqrt{2}$ in order to
match perturbative QCD in the $\alpha = 0$, the same can be done
here. In fact, doing this replacement for both quark and gluon
quasi-masses, the agreement with perturbative QCD is achieved.

The next natural question is: Does the scheme with $\alpha = 0$ also
reproduces lattice QCD data in the region near $T_c$? This question
is answered in the next section.

\section{Lattice data and qQGP}

In this section is discussed one of the most important features of
the quasi-particle approach: The fact that it is a useful tool to
describe lattice QCD data. Lattice QCD at non-vanishing chemical
potential uses a Taylor expansion in $(\mu/T)$ around $\mu=0$ to
compute observable quantities up to $\mathcal{O}(\mu^6)$
\cite{Allton:2005gk}. For the case of the pressure, the expression
is:

\begin{eqnarray}
\frac{P}{T^4}&=&c_0(T)+c_2(T)\frac{\mu^2}{T^2}+c_4(T)\frac{\mu^4}{T^4}+c_6(T)\frac{\mu^6}{T^6}+\cdots,
\label{taylor-p}
\end{eqnarray}
The coefficients $c_2,c_4,c_6$ have been computed in Ref.
\cite{Allton:2005gk} in the range $[0.76,1.98]$ for $T/T_c$. In the
present approach, the parameter $c_2$ will be used to constrain the
parameters $u$, $\lambda_s$, and $T_s$ appearing the running
coupling at finite $T$ and $\mu$, Eq. (\ref{g_s}). As seen in the
last section, the only case that reproduces perturbative QCD at is
for $\alpha = 0$. In that case,
$p=\frac{P}{T^4}=\frac{4}{\pi^2}[d_fI^T_5(r_f,a)+d_bJ_5(r_b)]$. On
the other hand, the definition of $c_2$ is
$c_2(T)=\frac{\del^2(P/T^4)}{\del a^2}\Big|_{a=0}$. Doing the
derivatives, one gets:

\begin{eqnarray}
\frac{\del^2{P}}{\del{a}^2}\Big|_{a=0}=\frac{1}{\pi^2}\left[d_f\left(\frac{\del}{\del{a}}\right)\left(G_3^T-r_f\frac{\del{r_f}}{\del{a}}I_3^T\right)-d_b\left(\frac{\del}{\del{a}}\right)\frac{\del{r_b}}{\del{a}}r_bJ_3\right]\Big|_{a=0}\nonumber\\
\frac{\del^2{P}}{\del{a}^2}\Big|_{a=0}=\frac{1}{\pi^2}\left[d_f\left(2I_3^T+\frac{r_f^2}{2}I_1^T-r_fI_3^T\frac{\del^2r_f}{\del{a}^2}\right)-d_br_bJ_3\frac{\del^2{r_b}}{\del{a}^2}\right]\Big|_{a=0}
\end{eqnarray}
The expressions for $I_1^T$, $I_3^T$, and $J_3$ have sums that have
to be stopped somewhere when doing the numerical calculation. The
criterion used was the radius of convergence of these sums, which is
$r_{b,f} < \pi$. With this in mind, one sees that for $r_{b,f}$
close to $\pi$, orders bigger that $k=5$ are negligible.

The data points from Ref. \cite{Allton:2005gk} where corrected by a
factor of 1.12, because as stated in Ref. \cite{Karsch:2000ps},
there should be a correction of 10-20\% when extrapolating the data
to the continuum. Following Letessier and Rafelski
\cite{Letessier:2003uj}, the first choice for the parameter $u$
appearing in the coupling was $u=1/\pi$. However, with this value,
the fit for the data below $T/T_c = 1.2$ is quite poor. If one
chooses, on the other hand, $u=1/2\pi$, agreement between lattice
and the calculation presented here is satisfactory, as long as $2.7
\leq \lambda_s \leq 3$ and $1.6 \leq T_s \leq 2$. With these results
the reduced pressure $\Delta
\frac{P}{T^4}=\frac{P(T,\mu)-P(T,0)}{T^4}$ can be calculated and
compared to the data. This is done in Fig. 1, with $\lambda_s = 3$
and $T_s = 1.98$. As seen, the agreement is quite good.


\section{Conclusion}

The number of different approaches that results in a consistent
thermodynamics of quasi-particles, whose masses are $T$ and $\mu$
dependent, is large, as the general formalism presented here shows.
The question is which one of the possible solutions is the most
adequate. Or, in other words, which one allows one to get the most
from it with some minimum input. Clearly, one can say that it is an
advantage to work with simpler expressions, preferably with
equations that allows one to make algebraic manipulations  all over
the calculation. And, of course, the adequate solution should be the
one that reproduce {\it both} perturbative QCD and lattice QCD.
Besides, it is desirable to have a solution with a field theory
analogue, as for instance the $\Phi$ derivable approximation, in
order to frame it in a stronger theoretical level.

From the solutions studied here, a whole class of them ($\alpha \neq
0$), which includes the Gorenstein-Yang type of solutions, have
problems when a perturbative expansion is made. Specifically,
although the thermodynamics consistency is achieved for the full
solution, it is lost order-by-order in the  perturbative expansion.
This implies that this class of solutions are not able to reproduce
perturbative QCD at finite $T$ and $\mu$ in any scenario. Of course,
when trying to reproduce lattice data near $T_c$, such problem does
not appear because, in that case, the full solution from the flow
equation for $B$ is used, and no order-by-order approximation is
made. Despite these problems, the solution GY1 has on its side the
fact that the entropy density and the particle number density
preserve the same form of the standard statistical mechanics,
exactly as in the $\Phi$ derivable approximation. On the other hand,
for the class of solutions with $\alpha = 0$ (pressure is unchanged
by the extra term $B$ while the other functions are modified),
order-by-order thermodynamics consistency in the asymptotic limit is
maintained and agreement with perturbative QCD is achieved.
Moreover, the lattice QCD data is reproduced down to $T/T_c \sim
1.02$.

In summary, a general consistent approach for the thermodynamics of
quasi-particles at finite $T$ and $\mu$ was presented for the first
time. It was show that the only class of solutions for the
thermodynamics functions capable to reproduce {\it both}
perturbative and lattice QCD is the one where the pressure is not
modified by the introduction of the extra term $B$.

This work was supported by FAPESP (04/15276-2) and CNPq
(307284/2006-9).

\begin{appendix}

\section{Appendix}
The necessary integrals that appear throughout this work are:
\label{App-A}
\subsection{Bose-Einstein Integrals}

The Bose-Einstein integrals are \cite{Gardim:2007ta,Haber:1981tr}

\begin{eqnarray}
J_n(r)=\frac{1}{\Gamma(n)}\int_{0}^{\infty}dx\frac{x^{n-1}}{(x^2+r^2)^{\frac{1}{2}}}\frac{1}{e^{(x^2+r^2)^{\frac{1}{2}}}-1},\hspace{2cm}n>0.\nonumber
\end{eqnarray}
The relevant integrals for our calculation, following
\cite{kapusta,jack,Gardim:2007ta} are:

\begin{eqnarray}
J_3(r)=\frac{\pi^2}{12}-\frac{\pi}{4}r-\frac{r^2}{8}\left(\ln\frac{r}{4\pi}+\gamma_E-\frac{1}{2}\right)-\frac{1}{4}\sum^{\infty}_{m=1}(m+2)j_mr^{2(m+1)}.\nonumber
\end{eqnarray}
\begin{eqnarray}
J_5(r)=\frac{\pi^4}{2^3425}-r^2
\frac{\pi^2}{96}+\frac{\pi}{48}r^3+\frac{r^4}{2^7}\left(\ln\frac{r}{4\pi}+\gamma_E-\frac{3}{4}\right)
+\frac{1}{32}\sum^{\infty}_{m=1}j_mr^{2(m+2)}.\nonumber
\end{eqnarray}
where
$j_m=\frac{(-1)^m\Gamma(m+1/2)\zeta(2m+1)}{2^{2m+1}\pi^{2m}\Gamma(1/2)\Gamma(m+2)}$.
The series are convergent for $r<2\pi$.

\subsection{Fermi-Dirac Integrals}

The Fermi-Dirac integrals are:

\begin{eqnarray}
I_n(r,\pm{a})&=&\frac{1}{\Gamma(n)}\int_{0}^{\infty}dx\frac{x^{n-1}}{\sqrt{x^2+r^2}}\frac{1}{e^{\sqrt{x^2+r^2}\pm{a}}+1},\hspace{2cm}n>0\label{I}\\
G_n(r,\pm{a})&=&\frac{1}{\Gamma(n)}\int_{0}^{\infty}dx\frac{x^{n-1}}{e^{\sqrt{x^2+r^2}\pm{a}}+1}.\hspace{3.8cm}n>0\label{G}
\end{eqnarray}

\ni As the system is composed by quarks and anti-quarks, the
thermodynamics functions can be written as a linear combination of
these integrals. Thus, the total integrals are:

\begin{eqnarray}
I^T_n(r,a)&=&I_n(r,a)+I_n(r,-a),\nonumber\\
G^T_n(r,a)&=&G_{n}(r,-a)-G_{n}(r,a).\nonumber
\end{eqnarray}

\ni These integrals satisfy the following recursion relations:

\begin{eqnarray}
\frac{\del{I^T_{n+1}(r,a)}}{\del{r}}&=&-\frac{r}{n}I^T_{n-1}(r,a),\label{ed-T-I}\\
\frac{\del{I^T_{n+1}(r,a)}}{\del{a}}&=&\frac{1}{n}G^T_{n-1}(r,a),\label{ed-T-G}\\
\frac{\del{G^T_{n+1}}(r,a)}{\del{r}}&=&-\frac{r}{n}G^T_{n-1}(r,a),\label{ed-T-G-r}\\
\frac{\del{G^T_{n+1}}(r,a)}{\del{a}}&=&nI^T_{n+1}(r,a)+\frac{r^2}{n}{I^T_{n-1}(r,a)}\label{ed-T-G-a},
\end{eqnarray}
with the initial conditions

\begin{eqnarray}
I^T_n(0)=\frac{2(1-2^{2-n})}{n-1}\zeta(n-1)\hspace{.6cm}n>2,\hspace{.3cm}\textrm{and}\hspace{.6cm}G^T_n(0)=0\hspace{.6cm}n>0
\label{I0}
\end{eqnarray}
where $\zeta(n)$ is
\begin{eqnarray}
\zeta(n)=\sum_{m=0}^{\infty}\frac{1}{n^x},\hspace{2cm}x>1,\nonumber
\end{eqnarray}
the Riemann zeta function. The procedure to compute the integrals
is inspired by Ref. \cite{Haber:1981tr}, where bosonic integrals
in the high temperature limit were computed.

If the integrals $I^T_1$ and $G^T_1$ are calculated, the recursion
relations can be used to determine the remaining integrals. For
the integral $I^T_1$ , the following identity is used
\cite{Haber:1981tr}:

\begin{eqnarray}
\tanh(z)=2\sum_{n=0}^{\infty}\frac{z}{z^2+[(n+\frac{1}{2})\pi]^2}=1-\frac{2}{e^{2z}+1}.\nonumber
\label{tanh}
\end{eqnarray}
Thus, the Fermi-Dirac distribution is written as:

\begin{eqnarray}
\frac{1}{e^{x}+1}=\frac{1}{2}-2\sum_{n=0}^{\infty}\frac{x}{x^2+[(2n+1)\pi]^2}.\label{f}\nonumber
\end{eqnarray}
Substituting this expression in $I^T_1$, with $x\rightarrow
(x^2+r^2)^{\frac{1}{2}}+a$, follows

\begin{eqnarray}
I^T_1(r,a)=\int_{0}^{\infty}dx\frac{1}{\sqrt{x^2+r^2}}\left[1+\right.\nonumber\\
\left.-2\sum_{n=1}^{\infty}\left(\frac{\sqrt{x^2+r^2}+a}{(\sqrt{x^2+r^2}+a)^2+[(2n+1)\pi]^2}+\frac{\sqrt{x^2+r^2}-a}{(\sqrt{x^2+r^2}-a)^2+[(2n+1)\pi]^2}\right)\right].\nonumber
\end{eqnarray}
Manipulating the integrand to simplify the integral, one gets:

\begin{eqnarray}
I^T_1(r,a)=\int_{0}^{\infty}dx\left[\frac{1}{\sqrt{x^2+r^2}}-4\sum_{n=0}^{\infty}\Re\left(\frac{1}{x^2+r^2-(a-(2n+1){\pi}i)^2}\right)\right]=i_1-4\sum_{n=0}^{\infty}i_{2,n}.
\label{I-T}
\end{eqnarray}

To solve the sub-integrals $i_1$ and $i_{2,n}$, it is necessary to
know the solution of two specifics integrals:

\begin{eqnarray}
z=\int_{0}^{\infty}dx\frac{x^{-\epsilon}}{\sqrt{x^2+b^2}}=\frac{b^{-\epsilon}}{2}B\left(\frac{1-\epsilon}{2},\frac{\epsilon}{2}\right)
\label{z-s},
\end{eqnarray}
\begin{eqnarray}
w=\int_{0}^{\infty}dx\frac{x^{-\epsilon}}{x^2+b^2}=\frac{1}{2b^{1+\epsilon}}B\left(\frac{\epsilon+1}{2},\frac{1-\epsilon}{2}\right).
\label{w-s}
\end{eqnarray}
where $B(p,q)$ is the Beta function. The factor $x^{-\epsilon}$
was introduced as a convergence factor. At the end of the
calculation, the limit $\epsilon\rightarrow 0$ is taken. Hence,
the sub-integrals are:

\begin{eqnarray}
i_1=\int_{0}^{\infty}dx\frac{x^{-\epsilon}}{\sqrt{x^2+r^2}}=\frac{r^{-\epsilon}}{2}B\left(\frac{1-\epsilon}{2},\frac{\epsilon}{2}\right).
\label{i1-solucao}
\end{eqnarray}
\begin{eqnarray}
i_{2,n}=\int_{0}^{\infty}dx\Re\left(\frac{x^{-\epsilon}}{x^2+r^2-(a-(2n+1){\pi}i)^2}\right)=\Re\left[\frac{B\left(\frac{1+\epsilon}{2},\frac{1-\epsilon}{2}\right)}{2[r^2-(a-(2n+1){\pi}i)^2]^{\frac{1+\epsilon}{2}}}\right].
\label{i_s2n-final}
\end{eqnarray}
Thus, the integral $I^T_1(r,a)$, Eq. (\ref{I-T}), is given by:

\begin{eqnarray}
I^T_1(r,a)=\frac{r^{-\epsilon}}{2}B\left(\frac{1-\epsilon}{2},\frac{\epsilon}{2}\right)-\frac{2\pi}{\sin[\frac{\pi}{2}(1-\epsilon)]}\sum^\infty_{n=0}\Re\left\{\frac{1}{[r^2-(a+(2n+1){\pi}i)^2]^{\frac{1+\epsilon}{2}}}\right\},
\label{I-1-solucao}
\end{eqnarray}
\ni For $\epsilon\rightarrow 0$, the functions can be expanded in
power series:

\begin{eqnarray}
b^{-\epsilon}=e^{-\epsilon\ln{b}}&=&1-\epsilon\ln{b}+\mathcal{O}(\epsilon^2),\nonumber\\
B\left(\frac{1-\epsilon}{2},\frac{\epsilon}{2}\right)&=&\frac{2}{\epsilon}+\ln4+\left(\frac{\pi^{2}}{6}+\ln^{2}2\right)\epsilon+\mathcal{O}(\epsilon^{2}),\nonumber\\
\csc\left[\frac{\pi}{2}(1-\epsilon)\right]&=&1+\mathcal{O}(\epsilon^2).
\label{expansions1}
\end{eqnarray}

\ni On the other hand, the denominator in the sum is rewritten as

\begin{eqnarray}
\frac{1}{[r^2-(a+(2n+1){\pi}i)^2]^{\frac{1+\epsilon}{2}}}=\frac{1}{[(2n+1)\pi]^{1+\epsilon}}+\hspace{7cm}\nonumber\\
+\frac{1}{[(2n+1)\pi]^{1+\epsilon}}\sum_{k=1}^{\infty}\frac{\Gamma(\frac{1+\epsilon}{2}+k)}{\Gamma(\frac{1+\epsilon}{2})\Gamma(k+1)}\left(\frac{a^2}{[(2n+1)\pi]^2}+\frac{2ai}{(2n+1)\pi}-\frac{r^2}{[(2n+1)\pi]^2}\right)^{k},\nonumber
\end{eqnarray}
where a Taylor expansion was done. This expansion is valid for two
cases: $a,r<\pi$ and
$|\left(\frac{a^2}{[(2n+1)\pi]^2}+\frac{2ai}{(2n+1)\pi}-\frac{r^2}{[(2n+1)\pi]^2}\right)|<1$.
Taking the limit $\epsilon\rightarrow 0$, one gets

\begin{eqnarray}
\sum_{n=0}^{\infty}\frac{1}{[(2n+1)\pi]^{1+\epsilon}}=\frac{2^{1+\epsilon}-1}{(2\pi)^{1+\epsilon}}\zeta(1+\epsilon)=\frac{1}{2\pi}\left(\frac{1}{\epsilon}+\gamma_E+\ln{\frac{2}{\pi}}\right),\nonumber
\label{zeta1}
\end{eqnarray}
where $\gamma_E\approx 0.5772$ is the Euler-Mascheroni constant.
The final result for $I^T_1$ is then

\begin{eqnarray}
I^T_1(r,a)=-\ln\frac{r}{\pi}-\gamma_E-\sum_{n=0}^\infty\frac{2}{2n+1}\sum_{k=1}^{\infty}\frac{\Gamma(\frac{1}{2}+k)}{\Gamma(\frac{1}{2})\Gamma(k+1)}\Re\left[\left(\frac{a^2-r^2}{[(2n+1)\pi]^2}+\frac{2ai}{(2n+1)\pi}\right)^{k}\right].
\label{I1T-solucao}
\end{eqnarray}
The integral $I^T_1$ was also computed in \cite{Schneider:2003uz},
using the Bose integrals of Ref. \cite{Haber:1981tr}. In
\cite{Schneider:2003uz}, however, the integral is truncated at
some order. In the present calculation, there is no truncation.

To compute the $I^T_3(r,a)$ integral, one needs $G^T_1(r,a)$.
Using Eq. (\ref{G})

\begin{eqnarray}
G^T_1(r,a)&=&G_1(r,-a)-G_1(r,a)\nonumber\\
&=&2\int_{0}^{\infty}dx\sum_{n=0}^{\infty}\left[\frac{\sqrt{x^2+r^2}+a}{(\sqrt{x^2+r^2}+a)^2+[(2n+1)\pi]^2}-\frac{\sqrt{x^2+r^2}-a}{(\sqrt{x^2+r^2}-a)^2+[(2n+1)\pi]^2}\right].\nonumber
\end{eqnarray}
After some manipulation,

\begin{eqnarray}
G^T_1(r,a)=-4\int_{0}^{\infty}dx\sum_{n=0}^{\infty}\Re\left[\frac{a-(2n+1){\pi}i}{x^2+r^2-(a-(2n+1){\pi}i)^2}\right]=-4\sum_{n=0}^{\infty}{i}_{3,n}.
\label{G-T}
\end{eqnarray}
The ${i}_{3,n}$ integral is similar to $i_{2,n}$:

\begin{eqnarray}
{i}_{3,n}=\Re\left\{\int_{0}^{\infty}dx\frac{x^{-\epsilon}(a-(2n+1){\pi}i)}{x^2+r^2-(a-(2n+1){\pi}i)^2}\right\}=\frac{B\left(\frac{1+\epsilon}{2},\frac{1-\epsilon}{2}\right)}{2}\Re\left\{\frac{a-(2n+1){\pi}i}{[r^2-(a-(2n+1){\pi}i)^2]^{\frac{1+\epsilon}{2}}}\right\}.\nonumber
\end{eqnarray}
As before, taking the limit $\epsilon\rightarrow 0$, one gets

\begin{eqnarray}
G^T_1(r,a)=a-\frac{7a}{2}\frac{a^2-r^2}{(2\pi)^2}\zeta(3)+\hspace{9cm}\nonumber\\
-2\sum_{n=0}^{\infty}\Re\left\{\frac{a+(2n+1){\pi}i}{2n+1}\sum_{k=2}^{\infty}\frac{\Gamma(\frac{1}{2}+k)}{\Gamma(\frac{1}{2})\Gamma(k+1)}\left(\frac{a^2-r^2}{[(2n+1)\pi]^2}+\frac{2ai}{(2n+1)\pi}\right)^{k}\right\}.
\label{G1T-solucao}
\end{eqnarray}
Notice that it was necessary to use both $k=0$ and $k=1$ terms in
the series expansion to get a finite integral. Using the recursion
relations (\ref{ed-T-I}) and (\ref{ed-T-G}), with Eqs.
(\ref{I1T-solucao}) and (\ref{G1T-solucao}), one can easily
integrate both expressions and use the initial condition Eq.
(\ref{I0}) to obtain $I_3^T$:

\begin{eqnarray}
I^T_{3}(r,a)=\frac{\pi^2}{12}+\frac{a^2}{4}-\frac{r^2}{4}\left(\frac{1}{2}-\gamma_E-\ln\frac{r}{\pi}\right)-\frac{7}{16}\frac{(r^2-a^2)^2}{(2\pi)^2}\zeta(3)+\hspace{3cm}\nonumber\\
-\sum_{n=0}^\infty\frac{\pi^2(2n+1)}{2}\sum_{k=2}^{\infty}\frac{\Gamma(\frac{1}{2}+k)}{\Gamma(\frac{1}{2})\Gamma(k+2)}\Re\left(\frac{a^2-r^2}{[(2n+1)\pi]^2}+\frac{2ai}{(2n+1)\pi}\right)^{k+1}.
\label{I3T}
\end{eqnarray}
To compute $G^T_3$  one needs Eqs. (\ref{ed-T-G-r}) and
(\ref{ed-T-G-a}). One starts taking the derivative of the solution
of Eq. (\ref{ed-T-G-r}) with respect to $a$. Then after some
manipulation of this result and comparing it to Eq.
(\ref{ed-T-G-a}), one gets

\begin{eqnarray}
G^T_{3}(r,a)=\frac{\pi^2a}{6}+\frac{a^3}{6}-\frac{ar^2}{4}-\frac{7\zeta(3)a}{16(2\pi)^2}{(r^2-a^2)^2}+\hspace{7cm}\nonumber\\
-\left[\frac{21\zeta(3)a}{16(2\pi)^2}{(4a^2r^2-r^4-3a^4)}-\frac{31\zeta(5)}{32(2\pi)^4}a{(r^2-a^2)^3}\right]+\hspace{4cm}\nonumber\\
-\sum_{n=0}^{\infty}[(2n+1)\pi]^2\Re\left\{\frac{a+(2n+1){\pi}i}{2(2n+1)}\sum_{k=3}^{\infty}\frac{\Gamma(\frac{1}{2}+k)}{\Gamma(\frac{1}{2})\Gamma(k+2)}\left(\frac{a^2-r^2}{[(2n+1)\pi]^2}+\frac{2ai}{(2n+1)\pi}\right)^{k+1}\right\}.
\label{G3T-solucao}
\end{eqnarray}
The second line is the contribution from $k=2$. Once $G_3^T$ is
known, $I_5^T$ can be determined. From Eqs. (\ref{ed-T-I}) and
(\ref{I3T}), and from Eqs. (\ref{ed-T-G}) and (\ref{G3T-solucao}),
one has:

\begin{eqnarray}
I^T_{5}(r,a)=\frac{7\pi^4}{2^545}-\frac{\pi^2(r^2-2a^2)}{96}+\frac{a^4}{96}-\frac{a^2r^2}{32}+\frac{r^4}{4^3}\left[\frac{3}{4}-\gamma_E-\ln\frac{r}{\pi}\right]+\frac{7\zeta(3)}{2^73(2\pi)^2}(r^2-a^2)^3+\nonumber\\
+\left[\frac{21\zeta(3)}{2^7(2\pi)^2}a^2(r^2-a^2)^2-\frac{31\zeta(5)}{2^{10}(2\pi)^4}(r^2-a^2)^4\right]+\hspace{4cm}\nonumber\\
\hspace{-2cm}-\sum_{n=0}^\infty\frac{\pi^4{(2n+1)^3}}{16}\sum_{k=2}^{\infty}\frac{\Gamma(\frac{1}{2}+k)}{\Gamma(\frac{1}{2})\Gamma(k+3)}\Re\left(\frac{a^2-r^2}{[(2n+1)\pi]^2}+\frac{2ai}{(2n+1)\pi}\right)^{k+2}.
\label{I5T-solucao}
\end{eqnarray}

\end{appendix}


{}
\newpage
\begin{figure}
\hspace{1.7cm}\includegraphics[width=12cm]{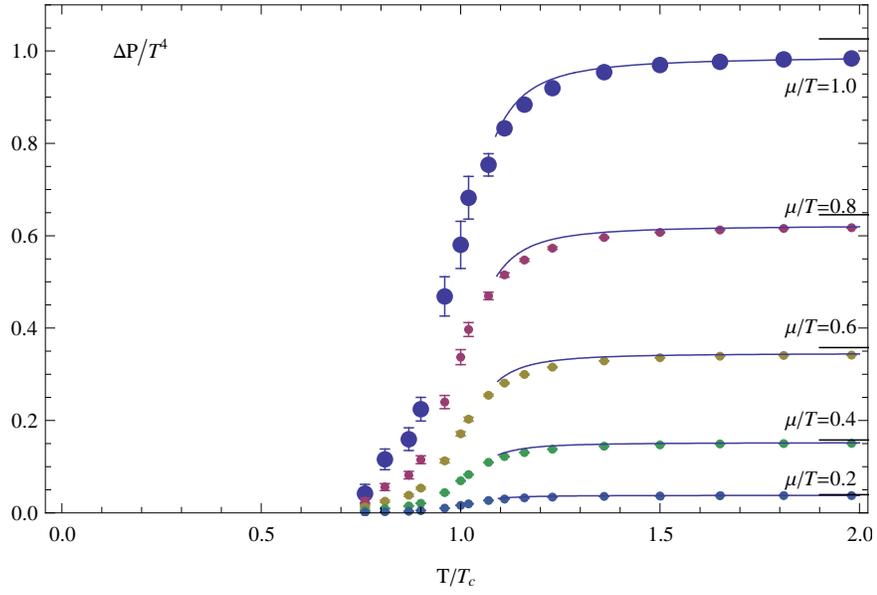} \caption{
The reduced pressure $\Delta p/T^4$. The points have been obtained
from lattice QCD \cite{Allton:2005gk}.} \label{fig}
\end{figure}
.
\newpage
\begin{table}[h]
\begin{tabular}{|c||c|c|c|c|c|}
  \hline\hline
  &$\Phi$ & S & \begin{tabular}{c}N\\ U \end{tabular}& \begin{tabular}{c} $\hat{N}_T$\\ $\hat{H}_T$ \end{tabular} & B \\
  \hline\hline
  $\alpha=0$ & $-T{\ln}Z_T$ & $-\langle\ln\hat{\rho}\rangle-\gamma\frac{B_S}{T}$ & \begin{tabular}{c}$\langle\hat{N}_T\rangle$\\ $\langle\hat{H}_T\rangle$\end{tabular} & \begin{tabular}{c}
  $\hat{N}-\frac{\lambda}{\mu}\hat{B}_N$\\ $\sum\hat{H}-\lambda\hat{B}_N-\gamma\hat{B}_S$ \end{tabular}& \begin{tabular}{c}
        $B_S=\frac{T}{\gamma}\sum\Big\langle\frac{\del{H}}{\del{T}}\Big\rangle$\\
        $B_N=\frac{\mu}{\lambda}\sum\Big\langle\frac{\del{H}}{\del{\mu}}\Big\rangle$
      \end{tabular}\\
\hline $\alpha\neq 0$ & $-T{\ln}Z_T+{\gamma}B$ &
$-\langle\ln\hat{\rho}\rangle-\gamma\frac{B}{T}$
&\begin{tabular}{c}$\langle\hat{N}_T\rangle$\\
$\langle\hat{H}_T\rangle$\end{tabular} &
\begin{tabular}{c}
 $\hat{N}-\frac{\lambda}{\mu}\hat{B}$\\ $\sum\hat{H}-\eta\hat{B}$
  \end{tabular} & $\frac{\del}{\del{m^2_i}}\frac{BT^{-\frac{\gamma}{\alpha}}}{\mu^{\frac{\lambda}{\alpha}}}=-\frac{T^{-\frac{\gamma}{\alpha}}}{\alpha\mu^{\frac{\lambda}{\alpha}}}\Big\langle\frac{\del{H_i}}{\del{m^2_i}}\Big\rangle$
     \\
\hline\hline
\end{tabular}
\caption{Summary of the two kinds of solutions for thermodynamics
relations.} \label{tabela}
\end{table}

\end{document}